# Anonymising Clinical Data for Secondary Use

**Authors:** Irene Ferreira[1], Chris Harbron[2,*], Alex Hughes[2], Tamsin Sargood[3], Christoph Gerlinger[4,5,*]

[1] Pharmaceutical Product Development (PPD). Granta Park, Great Abington, Cambridge, UK

[2] Roche Products, Welwyn Garden City, UK

[3] GSK, Stevenage, UK

[4] Statistics and Data Insights, Bayer AG, Berlin, Germany

[5] Department of Gynecology, Obstetrics and Reproductive Medicine, University Medical School of Saarland, Homburg/Saar, Germany

*Corresponding Authors

## Abstract

Secondary use of data already collected in clinical studies has become more and more popular in recent years, with the commitment of the pharmaceutical industry and many academic institutions in Europe and the US to provide access to their clinical trial data. Whilst this clearly provides societal benefit in helping to progress medical research, this has to be balanced against protection of subjects' privacy. There are two main scenarios for sharing subject data: within Clinical Study Reports and Individual Patient Level Data, and these scenarios have different associated risks and generally require different approaches. In any data sharing scenario, there is a trade-off between data utility and the risk of subject re-identification, and achieving this balance is key. Quantitative metrics can guide the amount of de-identification required and new technologies may also start to provide alternative ways to achieve the risk-utility balance.

Key words: Anonymisation, de-identification, risk, secondary use,

# INTRODUCTION

Secondary use of data already collected in clinical studies, understood here as the uses and disclosures that are different to the initial intended use for which the data was originally collected has become more and more popular in recent years, mainly due to transparency initiatives of regulatory authorities such as the European Medicines Agency's policy 0070 [1,2] and Health Canada's Public Release of Clinical Information (PRCI) policy[3] and the commitment of the pharmaceutical industry in Europe and the US to share their clinical trial data [4] (1)with independent researchers free of charge [5](1). Many academic institutions also provide access to their clinical trial data for secondary use [6](1). Although the initial intent of clinical data transparency was to allow the public to scrutinize drug regulatory decisions based on clinical trials sponsored by the pharmaceutical industry, the vast majority of use cases are to answer new questions with already existing data. For example, Mistry and her colleagues [7] performed a model-based analysis of the heterogeneity in the tumour size dynamics using the data of three different compounds from two sponsors via the clinicalstudydatarequest.com [5] (CSDR) (1)platform.

Secondary use of clinical study data has many methodological pitfalls [8], one if not the most important is the protection of the subjects' privacy. Health data is among the most sensitive personal data and is therefore especially protected data [9,10]. Typically, patient level data are anonymised before sharing. Anonymisation, as defined in detail below, is the process of changing the clinical trial (CT) data to protect the identities of individuals in the CT data by modifying their personal information to reduce the chances that a subject will be re-identified. However, the chance of re-identification can never be zero as long as some data is present, so anonymisation is also about making successful re-identification of a subject both difficult and costly. Moreover, it should be noted that this definition is relative to current technological capabilities and might change substantially over time. For example, a 15-year old boy was able to re-identify his anonymous sperm donor father using a publicly available DNA database designed for genealogic research, a technology not available at the time of the sperm donation [11].

This paper focuses on the use of various anonymisation techniques and especially on techniques to assess the "quality" of an anonymisation. We will especially focus on two scenarios:
(1) clinical study reports (CSR) published in redacted form e.g. on the EMA website for unrestricted download as part of Policy 0070 [1] or PRCI [3]
(2) individual patient or participant data (IPD) that are made accessible under controlled conditions e.g. via secure portals [5,6]. IPD is the clinical study person-specific data recorded separately for each participant.

In section 2 we discuss some current practices for the anonymisation of CSRs and IPD and in section 3 we describe some of the associated statistical challenges along with some alternative approaches. Section 4 illustrates some of these concepts through case studies and section 5 discusses some of the current and future challenges surrounding data sharing. We also provide some key definitions related to anonymization as an Appendix.

# CURRENT PRACTICES

## CLINICAL STUDY REPORTS

Both the EMA and Health Canada now require pharmaceutical companies to make clinical documents available on their websites. Both require acceptance of terms of use (ToU) prior to viewing/downloading the documents. In addition, to access the EMA website, an individual must create an EMA account which requires the registration of an email address. The ToU includes a commitment not to re-identify individuals. However, it is not possible to limit access to the clinical documents, they are available to any member of the general public, provided they agree to the ToU. Clinical documents are rich in IPD and must be anonymised to account for the significantly increased risk of re-identification.

There are two different approaches to anonymization that can be seen when browsing examples on the EMA's Clinical Document Portal and Health Canada's PRCI portal:

1- Redaction only
2- Anonymization that includes transformation, generalizing and redaction

When selecting approach sponsors are asked to consider how they protect privacy whilst retaining a level of clinical utility in the documents. This is complex and the decision made is often based on the capabilities a sponsor has and the tools available to them.

Redaction only: The most common approach used to date by sponsors who are publishing anonymized or redacted clinical packages is the use of pure redaction. Redaction is the blacking out personal information found in the text on all identifiers such as names, signatures, contact details, subject ID numbers, dates, all free-text, names of investigators, site staff, research institutions and staff, vendor and labs and sponsor company staff.  Reliance on only redaction limits clinical utility of the published documents. For example, for subjects mentioned repeatedly across the document, the connection is lost when relying fully on redaction.

Anonymization: Some sponsors are starting to anonymise the CSRs using a variety of anonymisation techniques that go beyond redaction. These techniques eliminate the ability to connect an individual to the information but allow for a greater level of clinical utility for those using the documents for additional research. Some examples of some common anonymization techniques include:

- Recoding of subject identifiers to create new unique identifiers that can be retained throughout the document
- Applying an offset to dates or replacing dates with relative day thus retaining a level of clinical utility to understand the relationship between key treatment dates and events for a given patient

- Grouping of indirect identifiers such as combining all individuals from countries in EU into one group or creating age bands instead of redacting age all together. These groups again allow some information to be retained to provide utility when information is being read by others.
- Redaction is also used in many cases as well for things such as contact details, signatures, or sensitive personal information in one's medical history (e.g. Suicide, STDs, conditions of mental health, visible scaring, etc.)

Sponsors must take special care and consideration when the same data is presented in various summary level tables. It can become obvious in the case of small numbers such as zeros and ones, the collective view of one patient's demographics making it then easier to connect that one information to other information that may be in the public domain or known in another way. In addition to looking at patients individually to protect them, sponsors must also consider the information that can be collected about individuals through aggregated data tales

## INDIVIDUAL PATIENT DATA

Anonymised IPD is being made available to external research teams for secondary use via controlled environments such as Vivli [6] and CSDR [5] where substantial technical and organisational measures are in place to reduce the risk of re-identification. The external research team must submit a research proposal which is assessed for scientific merit by an independent research panel and for limitations of the subjects' informed consent by the sponsor. If the proposal is approved and the proposed research respects the subjects' informed consent, the research team must sign a legally binding data sharing/use agreement which includes a commitment not to attempt to re-identify individuals. The IPD is shared via a secure environment which does not allow the download of the IPD and access is limited to the researchers named on the research proposal.

Pharmaceutical companies also perform internal secondary analyses of IPD. These analyses must respect the terms of the informed consent under which the data was collected from subjects. An additional consideration is that, under the General Data Protection Regulation (GDPR), a study participant may object to the processing of their IPD at any time, including after the clinical trial has concluded.  It may only relate to a particular type of processing (e.g. internal or external secondary analyses, anonymisation). GDPR does not apply to personal data that has been anonymised, "…The principles of data protection should therefore not apply to anonymous information, namely information which does not relate to an identified or identifiable natural person or to personal data rendered anonymous in such a manner that the data subject is not or no longer identifiable." This Regulation does not therefore concern the processing of such anonymous information, including for statistical or research purposes." As such by anonymising the IPD on completion of the trial, the risk of a participant objecting to further processing of their IHD is minimised and, as the anonymised IPD can no longer be linked to an individual, the full anonymised dataset can continue to be used for secondary analyses. Access to the anonymised IPD and pseudonymised IPD must be controlled to minimise the risk of re-identification and written standards should be implemented that require internal research teams not to attempt to re-identify individuals. Technical and organisational measures can be implemented to reduce the risk of re-identification.

## The Risk-Utility Trade-Off

The key trade-off in anonymising personal data lies between the protection of subject confidentiality and data utility. Clearly a decision not to share any data at all absolutely minimises the risk of breaching subject confidentiality but has no data utility. On the other hand, sharing the entirety of the data from a clinical trial will maximise the utility, but has a high risk of allowing subject re-identification. Both dimensions of this trade-off are highly scenario dependent and require understanding both the data as well as the surrounding context.

Researchers and the agencies want sponsors to publish Clinical Study Reports with the greatest level of utility. In the end, the sponsor has to use an appropriate risk measurement metric, such as k-anonymity, to determine what information is safe to disclose and what is not. Figure 1 proivides an illustration of how variation in the anonymization techniques selected deliver different levels of privacy protection vs. clinical utility.

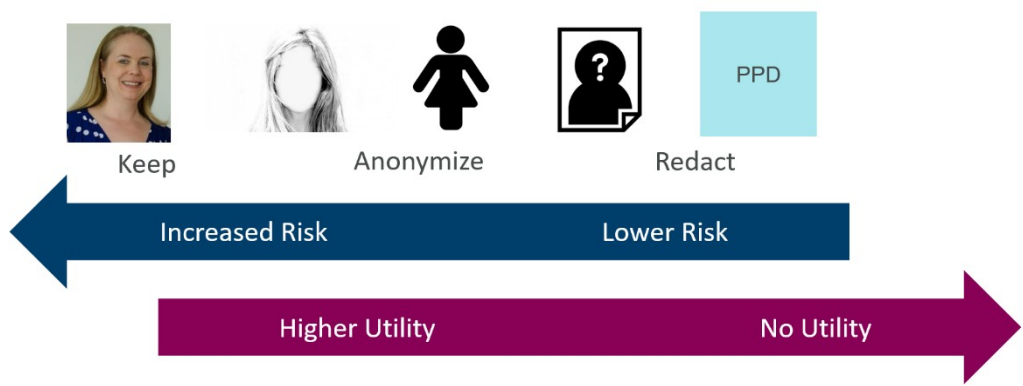

*Figure 1: The Risk-Utility Trade-Off*

# Re-identification Attacks

Elliot et al [17] emphasise that a well-informed decision about whether and what anonymisation is needed requires consideration of both the data and their environment in what they term the data situation. As well as understanding the properties of the data this also requires understanding a range of factors including legal and ethical considerations the motivation and capabilities of any potential attackers and the potential impact of any disclosure.

From this understanding different types of attack can be identified, two key dimensions of which are the motivation and the risk type.  The potential motivations of an attacker are many-fold, and two extremes are the "nosy neighbour" attack in which the attacker targets finding out information about a specific individual or the "demonstration attack" in which the attacker targets finding out information about any individual with the motivation of demonstrating a weakness in the system. If all of the identifiers upon which any match would be made are publicly available then the 'demonstration attack' has a higher risk of re-identification than a nosy neighbour attack as the nosy neighbour attack is seeking information about a specific individual whilst the demonstration attack is successful in the easier challenge of identifying information about any individual. If however the identifiers include information which may only be reliably known by close associates of an individual, then this may decrease the likelihood of success for a demonstration attack and make a nosy neighbour attack, where more reliable and detailed information may be known about the subject, the higher risk.

Examples along the risk type dimension are "prosecutor risk" where the individual is known to be in the trial, for example through self-disclosure in conversation or on social media, or "journalist risk" where the individual is just known to be in the wider population, for example to have the indication in which the study is being run, but may or may not be part of the study.

Many of the metrics around calculating risk are based on the concept of an equivalence class, that is a set of subjects with the same set of identifying variables or quasi-identifiers whom it would not be possible to distinguish between based upon those variables. The set of equivalence classes partitions the study population into $n$ distinct sets. Note that the equivalence classes can potentially vary from attacker to attacker depending upon what information or quasi-identifiers on subjects they have access to.

From the partition of the study population into equivalence classes  a number of risk metrics can be calculated [18]:

|   | Description | Formula |
|---|---|---|
| 1 | Proportion of subjects in an equivalence class smaller than a threshold size | $\dfrac{\sum f_i \times I(f_i < \tau)}{n}$ |

| 2 | Average size of equivalence classes | $\frac{|J|}{n} = \frac{\sum f_i}{n}$ |
|---|---|---|
| 3 | 1 / Size of smallest equivalence class | $\frac{1}{min(f_i)}$ |

Where $f_i$, i=1,...,$n$ are the sizes of the set of the n equivalence classes, J is the total number of subjects and $\tau$ is a threshold.

When we are considering a demonstration attack, then the third risk metric is the most relevant and is the most commonly quoted and used. This is the concept of k-anonymity where if satisfied every individual is in an equivalence class of at least size k, i.e. given a set of identifying variables these would refer to at least k individuals. A commonly quoted threshold for open access public data sharing for k-anonymity, including by the EMA [11] and Health Canada [12] is 0.09, equating to a minimum equivalence class size of 11. In scenarios where greater controls are in place, such as using a controlled environment with a Data Use Agreement in place then a higher risk threshold corresponding to a smaller minimum equivalence class size may be acceptable.

If a demonstration attack was considered unlikely, so that any attack would be focussed on specific individuals, then the first two metrics become relevant, as the demonstration attack requires reducing the risk of re-identification for all subjects, whilst an attack targeted at specific individual only requires ensuring the risk of re-identification for that individual is low, and if a small number of individuals in a large dataset are at a higher risk of re-identification then it is unlikely that one of those will be the targeted individual. The first metric assesses the proportion of individuals who are at an increased risk of re-identification through being in a small equivalence class, compared to using the third metric which requires no individuals to be in a small equivalence class. However using this first metric requires considering two parameters or thresholds, the desired size of the equivalence class and the proportion of individuals falling into a smaller equivalence class.

The strict average metric has been proposed [19] which is a combination of second and third risk metrics, giving protection for both types of attack by requiring both no subjects to have unique records or be in an equivalence class of size 2 as well as the average equivalence class size to be at least 10, i.e. metric 3 is less than 0.5 and metric 2 is less than 0.1.

An example of the use of these three metrics to guide decision making is shown in the Individual Patient Level Data Example.

Two key data de-identification strategies are typically deployed in combination to reduce the level of risk to acceptable levels are removing, where variables are omitted from the data set, and grouping. In grouping, similar values taken by variables are grouped together for example moving from a variable of country to a continent variable, or in the case of a numerical variable such as age, grouping values into age bands. The impact of both these strategies is to merge adjacent equivalence classes creating fewer and larger equivalence classes.

A feature of k-anonymity is that the number of equivalence class grows exponentially with the number of variables which are direct-identifiers or quasi-identifiers, and the size of the smallest equivalence class shows a corresponding rate of decrease with increasing numbers of identifying variables. This means that for a typically sized clinical trial, only a limited number of quasi-identifiers can be shared whilst still meeting the risk metric threshold. Note that many variables within a clinical trial, for example the values of lab tests, will not be identifiers and so will not impact the calculation of k-anonymity.

k-anonymity has the limitation that it may still allow inference on sensitive attributes even within a large equivalence class if all or most of the members have the same value for the sensitive attribute. l-diversity is an extension of k-anonymity additionally requiring that any equivalence class contains l well represented values for the sensitive attribute. The most common application of l-diversity is with a binary sensitive attribute that can take two values and a value of l=2, which requires that every equivalence class contains examples of both levels of the sensitive attribute or alternatively that in a frequency tabulation of equivalence class against the sensitive attribute no cells would have a frequency count of zero. Failure to achieve 2-diversity would mean that knowing the equivalence class of some individuals would allow the value of the sensitive attribute to be inferred.

Whilst 2-diversity will protect against certain identification of an individual's sensitive characteristic based upon their quasi-identifiers, it may still allow probabilistic statements to be made if the distribution of the sensitive attribute within an equivalence class is different to the wider population. For example if a rare sensitive condition is only present in 10% of the population, but 90% of an equivalence class exhibit this condition then knowing that an individual is in the equivalence class considerably increases the likelihood that the individual has that condition. t-closeness addressed this risk by requiring that the distribution of a sensitive attribute within an anonymised group should not differ from the global distribution by more than a threshold t, measured using an appropriate distance metric. The implementation of t-closeness will depend upon the type and distribution of the sensitive variable. One option may be requiring for all equivalence classes that the statistical test comparing the distribution within the equivalence class with the combination of all other equivalence classes is non-significant.

As we move from k-anonymity to l-diversity to t-closeness, we are applying increasingly stringent requirements. This will result in a decreased risk of re-identification or disclosure of sensitive information for a subject, but also a corresponding reduction in the data utility as more information is lost from the dataset.

# Risk of Re-Identification

It can be informative to decompose the risk of re-identification into two terms, both of which have the potential to be controlled by different types of measures.

Probability of Re-Identification = Probability(Attack) x Probability(Re-Identification / Attack)     (1)

Further granularity and insight may be achieved through decomposing this equation by considering the different kinds of attack that can occur within any particular scenario:

$$\text{Probability of Re-Identification} = \sum_{\text{Types of Attack}} \text{Prob}(Attack_i) \times \text{Prob}(\text{Re-identification}/Attack_i)$$

In a non-public sharing scenario, the Institute of Medicine report on sharing clinical trial data [19] identifies three types of attacks:

- a deliberate re-identification attempt by the data recipient,
- an inadvertent (or spontaneous) re-identification by the data recipient
- a data breach exposing the data to a wider audience.

These could potentially be decomposed further, for example by distinguishing a deliberate re-identification attempt made with respect to a single individual (a nosy neighbour attack) and a deliberate attempt to identify any individual within the data (a demonstration attack).

For re-identification to occur, both an attack has to occur, and the attack has to be successful. This gives rise to two main strategies for controlling the risk of re-identification which both operate by reducing one of the terms in (1) to close to zero, thus controlling the overall probability of re-identification.

On controlled data sharing platforms e.g. Vivli or CSDR, risk is reduced to an acceptable level by minimising the first term of the equation, the probability of an attack occurring, through the platforms' tracking capabilities and the legal and reputational implications of contravening the agreement that any requestor has to sign up to. If these controls are adequate, which will minimise the likelihood of an attack, then we do not have to be concerned about the likelihood an attack will be successful and only basic de-identification processes, removing or pseudoanonymising direct identifiers, may be required, thus maximising the data utility. Taking a more detailed view by considering the different types of attacks listed above, a data sharing platform may keep the overall probability of an attack low by keeping the probabilities of all of these types of attack low.

Requiring signature of a legally binding data sharing agreement will constrain researchers' behaviours to not commit a deliberate re-identification, requiring researchers to access any shared data within a secure data access system will minimise the chances of a data breach and the probability of inadvertent re-identification can be lowered by removing any obvious identifiers from the data. In most circumstances inadvertent re-identification is highly unlikely as the analyst will only know a small fraction of people in the total population, and the study will only contain a small fraction of people in the total population and so it is unlikely these will intersect, particularly if the data comes from an international study. There may be specific circumstances which will require special attention for example if the study contains well-known subjects in the public eye or in rare diseases with a strong patient community if the analyst is a member of member of that community. Internal data sharing within a company also generally falls into this category where, as an employee, the requestor will have signed up to follow the company's code of conduct which will include appropriate behaviours around respecting personal data and the company will have stringent data protection policies in place to prevent any data breaches.

On open data sharing platforms, e.g. GAAIN [20], as the data is effectively open to anyone with any level of motivation or capability of carrying out a successful attack, we have to assume that the probability of an attack in (1) is 1. Therefore, in order to control the overall probability of re-identification requires adapting the data to reduce the risk of re-identification in the event of an attack to be close to zero, which is likely to incur a loss of data utility.

A consequence of this is that data shared through data sharing platforms under a data sharing agreement, is likely to have greater utility to any requestor, than data which is obtained from open data sharing platforms.

Hybrid strategies, assuming that both terms in (1) are low so that their product is low enough to be an acceptable risk have been proposed but have the potential to be quite non-robust as they require numerous assumptions to estimate both terms giving a false sense of confidence or accuracy, in contrast to the simpler strategies of just reducing one or other of the terms to close to zero. The challenge of this approach is obtaining robust values for both terms of the equation.

The first term in the equation, the probability of an attack, has been estimated from historical precedent. This has the risk that this may not be constant over time as the external environment and actors evolve or may not translate to new scenarios such as EMA Policy 70 or Heath Canada's PRCI.

For the second term in the equation, the probability of re-identification given an attack, some publications have quoted a relationship that the probability of re-identification of a subject in an attack is the reciprocal of the size of that subjects' equivalence class. This relationship is true with an equivalence class of size one, as a unique match can be identified with probability one giving certain re-identification. However, more generally with larger equivalence classes this relationship is measuring the probability of a correct re-identification assuming the behaviour that after identifying the equivalence class with the same attributes as the individual they are wanting to identify in the data, the attacker makes a random selection from the equivalence class and claims this is an accurate re-identification. This is the probability of an event assuming a specific and unlikely behaviour by the attacker. In practice an attacker faced with an equivalence class containing a number of matches may be likely to conclude they cannot make a reliable identification. All that could be deduced is that there is a chance of one in the size of the equivalence class that the individual has the specific confidential attribute. As this may be combined by an attacker with other external information there is a driver to make a higher minimum requirement for an equivalence class size beyond just avoiding uniquely described individuals.

As a simple hypothetical example consider the following two scenarios where calculations could give the same result but which carry very different levels of risk. Suppose a 48-year-old Swedish male subject is known to be in a particular trial, and knowledge of his age, gender and nationality is publically known. There are 10 48 year old males in the study, but he is the only Swede. As part of the trial protocol he has been genotyped and found to have a deleterious mutation, but has requested not to be informed of the outcome of the genetic testing which should remain confidential.

In scenario A the age, gender and mutation status of all individuals will be made freely available on the internet.  Here we assume that if the data is freely publically available there will be an attack, i.e.  a probability of attack of 1, and we observe an equivalence class size of 10 based on the identifiers of age and gender. We are interested in preventing an attack which would disclose sensitive genetic information based on commonly known individual identifiers.

In scenario B, a random process will occur with a probability of 0.1 that the age, country and mutation status of all individuals will be made freely available on the internet. Again assuming that if the data is freely publically available there will be an attack this gives a probability of an attack of 0.1, the probability of publication, and an equivalence class size of 1 based on the identifiers of age, gender and nationality giving a certainty that if an attack occurs it will be successful.

Applying equation (1) in scenario B, the probability of identification and that the individual's sensitive genetic result will be disclosed with certainty is 1 x 0.1 = 0.1. In scenario A, although the calculation 0.1 x 1 = 0.1 could be performed to give the same value, this is highly different situation as there is no chance of being able to make any certain deduction about the individual, which is a preferable scenario.

There is still is the risk that an attacker might be able to make inferences for example when all members of the equivalence class share a common attribute or attributes, which is the motivation for l-diversity discussed previously.

## Data Utility

The utility of data is highly dependent upon how the data is going to be used. If an analysis can be well defined in advance, then simple steps such as removing all variables not playing a part in the analysis do not impact the data utility, whilst decreasing any subject re-identification risk. Understanding the purpose of analysis would also allow sponsors to verify that any anonymisation process does not materially affect the conclusions from any analysis of the data before releasing it to the requestor.

Where, in contrast, as is often the case, the analysis is to be more exploratory in nature, maybe even using techniques such as machine learning, then the utility is harder to define and guarantee. In these situations, a pragmatic approach of applying minimal anonymisation in order to satisfy the required risk re-identification metrics may be the most pragmatic approach. These decisions could additionally be guided by subject matter expert knowledge for the indication under study to prioritise those variables most likely to be informative or of interest to a requestor. As long as it does not create unacceptable reidentification risks consideration should also be given to providing structure within the data which will support researchers, for example with adverse events also sharing anonymised subject IDs and relative dates in order to allow deeper understanding of individuals' time courses in safety profiles.

This is even more of a challenge for data sharing platforms which allow for an open release of data to any requestor, as these must aim to provide data with utility for any possible question which could be asked of the data, as opposed to those platforms (e.g. CSDR) which release data in response to specific research proposals.

## Alternatives To Anonymisation

New methodological developments are opening up the potential for alternative approaches to anonymisation to allow secondary use of data without the risk of patient re-identification. These are emerging areas where the level of adoption and utility to researchers is still evolving.

### FEDERATED COMPUTING
Ideas from distributed or federated computing [21] may also be helpful in some situations, for example IPD meta-analyses. Rather than the analyst requiring all of the data from multiple sources to be shared with them, taking the data to the analyst, an alternative approach is to take the analysis to the data, so that the individual level patient data remains at sites. As a simple example, if an analysis can be pre-specified, instead of each site having to perform anonymization to share their data, they could instead perform the pre-specified analyses on their data and just share a set of summary statistics with the central analyst who can then combine these into an approximation to the true IPD level analysis, i.e. the analysis that would have been performed if all the IPD was available to the analyst.

## USE OF SYNTHETIC DATA

Synthetic data is an alternative approach for scientific data sharing, circumventing the need to share the data from any individual subjects and instead sharing synthetically generated data which does not relate to individual subjects but maintains the structure and relationships from the original data set. James et al [23] discuss a range of use cases for using synthetic data and show that the requirements for how closely the synthetic data needs to capture the full structure of the original data is dependent upon the use case.

If synthetic data can be generated with properties that make it equivalent to analysing the actual data for the desired analysis, sharing this synthetic data may be an alternative to anonymisation, as no data from real individuals is shared. However care still needs to be taken to ensure that inferences cannot be made from the synthetic data which may re-identify patients from the original dataset from which the synthetic data was derived.

## NLP

Developments in Natural Language Processing (NLP) have the potential to be able to detect and address identifiers or quasi-identifiers within Clinical Study Reports, which would substantially reduce manual efforts. The adoption challenge for using these tools is to be able to ensure high levels of quality which will generally currently still require significant to full manual quality control.

# CASE STUDIES

## Example 1 – Clinical Study Report

A Clinical Study Report will have two key areas requiring care to avoid re-identification risks: text, in particular narratives giving a description of adverse events within the study and data tabulations with small cell counts.

A typical unanonymsed narrative describing an adverse event may read:

> *Subject '000478' male, aged 35, from Argentina, re-started IP after recovery from the traffic accident on 16/Oct/2006 and developed a rash on his face on 17/Oct/2006. IP was stopped on 01/Nov/2006*

This contains multiple personal identifiers which especially when combined with knowledge of the patient and potentially other text or tables within the Clinical Study Report may create a risk of identification of the individual.

The simplest approach would be to redact all personal identifiers:

> *Subject '██████████' ████, aged ████, from ██████████, re-started IP after recovery from the traffic accident on ██████████ and developed a ████ on his face on ██████████. IP was stopped on ██████████*

This has removed identifiers of the patient's subject number, gender age and location as well as details of the adverse event, its duration and its impact on the patient's treatment. Whilst this has reduced the risk of re-identification, the utility of the resulting text for understanding the impact of treatment is minimal.

A more fruitful approach may be anonymization beyond redaction:

> *Subject '000798' ████, aged 30-40, from South America, re-started IP on 15/Sep/2005 and developed a [Skin and subcutaneous tissue disorders] on ████ ████ on 16/Sep/2005. IP was stopped on 02/Oct/2005*

Some identifying information such as the patient's gender and the location of the adverse event is still redacted. However, wherever possible some level of information is retained. Dates are consistently offset so that the duration and sequencing of events is maintained, although the actual dates are no longer correct and so cannot be used to make an identification of an individual. Other variables are grouped  This still maintains much of the utility of the original text without disclosing some of the more identifying information which often adds little extra to the clinical understanding of the patient's experience.

As with individual patient data, metrics such as k-anonymity can guide the level of grouping required to reduce the risk of re-identification to an acceptable level.

Anonymization beyond redaction clearly allows for the greatest utility. However, it is the most time consuming and resource burdensome approach. Notably it is also limited in some situations around what information can be disclosed

Data tabulations can also create a risk of re-identification, typically when there are cell counts of zero or one.

In this tabulation of age group against a history of a certain medical condition (the details of which are withheld here to maintain confidentiality) there is a count of zero in the cell for no history in the 40-45 age group.

|  |  | Age Group | | | | | | |
| --- | --- | --- | --- | --- | --- | --- | --- | --- |
|  |  | 40-45 | 45-50 | 50-55 | 55-60 | 60-65 | 65-70 | >70 |
| History of specific medical condition | No | 0 | 5 | 13 | 25 | 33 | 14 | 16 |
|  | Yes | 2 | 4 | 9 | 16 | 21 | 8 | 11 |

This allows the reader to infer that both of the 40-45 year olds had a history of this condition, sharing these individuals' medical history to anyone who knows they participated in this study. Complete redaction of the table was a possibility but the decision was made to use wider age categories to group the patients as shown below in order to reduce the risk of re-identification whilst maintaining the utility to be able to explore the relationship of the medical history and age.

|  |  | Age Group | | | |
| --- | --- | --- | --- | --- | --- |
|  |  | 40-50 | 50-60 | 60-70 | >70 |
| History of specific medical condition | No | 5 | 38 | 47 | 16 |
|  | Yes | 6 | 25 | 29 | 11 |

## Example 2- Individual Patient Data:

The Global Alzheimer's Association Interactive Network (GAAIN [23]) provides a federated data platform designed to foster cohort discovery, collaboration and sharing, currently containing IPD from 51 data partners and freely accessible via a rapid registration. The Alzheimer's Prevention Initiative (API) Colombia Trial is a study evaluating anti-amyloid antibody crenezumab versus placebo treatment in cognitively unimpaired PSEN1 E280A mutation carriers, also including placebo-treated non-carriers. As this was a unique subject cohort in a disease with high unmet medical need, there was interest in making baseline data from this study available to the scientific community through the GAAIN platform, although information on a genetic mutation which is highly predictive for the development of Alzheimer's made this data particularly sensitive.

As there were a limited number of older patients in the cohort, and a correlation of age with a lower risk of carrying the genetic mutation would be expected, any presentation including age and mutation status was considered a high risk of disclosing sensitive information around mutation status. So, as an initial precautionary step, all 10 subjects, including both carriers and non-carriers, in the cohort aged over 54 were removed from all presentations of results reducing the number of patients from 252 to 242.

The data initially shared via GAAIN was purely baseline values and contained 8 variables

- Subject ID {randomly assigned already scrambled}
- Gender {"M", "F"}
- Race {"OTHER"}
- Age {30, ..., 60}
- Ethnicity {"HISPANIC OR LATINO"}
- Country {"Colombia"}
- CDR Global {0,0.5}
- MMSE Total Score {23,...,30}

Three of these variables, Race, Ethnicity and Country, are identical for every subject and given the publicly available trial set up would already be known and the Subject ID had already been scrambled by randomly assigning a new value to break the link with the original data, so was effectively an arbitrary index number. CRD and MMSE are two assessments of cognitive performance used to diagnose dementia and assess its severity which would be considered medically confidential information. The two remaining variables, gender and age, were considered identifiers

Table 2 shows the series of steps that we took to achieve an adequate level of anonymisation whilst still maintaining as high a data utility as possible

| Step | Scenario | Proportion of subjects in an equivalence class smaller than a threshold size (5) | 1 / Average size of equivalence classes | 1 / Size of smallest equivalence class |
|---|---|---|---|---|
| 1 | Raw Data | 0.115 (28/242) | 0.20 (48/242) | 1 |
| 2 | Removing the variable gender from the dataset | 0.05 (12/242) | 0.10 (25/242) | 1 |
| 3 | Grouping age into age bands of 10 years. Retaining gender with the dataset | 0.017 (4/242) | 0.024 (6/242) | 0.25 (1/4) |
| 4 | Grouping age into age bands of 10 years. Omitting gender from the dataset | 0 | 0.012 (3/242) | 0.1 (1/10) |
| 5 | Grouping age into 4 age bands 30-34, 35-39, 40-44, >=45. Omitting gender from the dataset | 0 | 0.017 (4/242) | 0.03 (1/30) |

| 6 | Grouping age into 4 age bands 30-34, 35-39, 40-44, >=45. Retaining gender with the dataset | 0 | 0.033 (8/242) | 0.06 (1/18) |

1. The raw data gave some subjects with unique observations giving equivalence classes of size 1, i.e. unique values of age and sex within the dataset and so we recognised there was a need for anonymization.
2. Age was considered a more important variable than gender within this dataset, so we tried removing gender from the dataset just leaving age, but we were still left with subjects with unique ages who could be directly identified.
3. We concluded that we would need to group age into age bands in order to achieve acceptable risk metrics, so we re-introduced gender into the dataset and grouped age into age bands of 10 years. This was still giving some unacceptably small equivalence classes, the smallest being 4 subjects, so we recognised further anonymization steps were necessary.
4. We then removed the gender variable from the dataset. This gave acceptable risk metrics, but the data utility was limited with subjects just being grouped into one of 3 wide age bands.
5. We then examined the distribution of ages, and we realised that we could both increase the utility of the data and obtain a better distribution of equivalence class sizes by changing the age bands from 30-39, 40-49, 50-59 to 30-34, 35-39, 40-44, >=45
6. With this more even distribution of ages giving acceptable risk metrics, we then investigated further increasing the data utility by reintroducing gender into the dataset. With this anonymised dataset, the smallest equivalence class size was 18 and so this was the data shared on the GAAIN platform alongside non-identifying data such as the results of clinical imaging. This final tabulation of patients by age-band and gender shared with GAAIN is shown in table 3

| Gender | Age Band | | | | |
| --- | --- | --- | --- | --- | --- |
| | 30-34 | 35-39 | 40-44 | >=45 | Total |
| F | 55 | 36 | 30 | 30 | 151 |
| M | 23 | 30 | 20 | 18 | 91 |
| Total | 78 | 66 | 50 | 48 | 242 |

## Discussion

There is general recognition that data sharing and data protection are objectives that need to be achieved at the same time. Unfortunately there is a trade-off of these objectives as more data sharing increases the risk to subjects' privacy and more data protection reduces the utility of the data shared.

We have discussed two levels of data sharing; CSRs and IPD. IPD, anonymised by taking into account the proposed analysis and shared under controlled conditions, will provide much higher levels of data utility than data shared in an unrestricted manner where the degree of generalisation or redaction required to still maintain subject confidentiality will impact the data utility. Therefore, for many analyses IPD shared under an agreement will be preferred.

In this paper we have shared the current thinking around data sharing and anonymisation. However, this is an evolving area with on-going research and so it is possible that new technologies may emerge which will change the paradigm, for example adopting ideas from synthetic data, federated computing or cryptography.

However there are remaining questions that will continue to be discussed. If we consider demonstration attacks which may lead to reputational risk, this may lead us to consider multiplicity. Should we consider the likelihood for re-identification occurring on a study-by-study basis? Or, as a large pharmaceutical company typically has hundreds of studies on-going and even a small risk for each study could add up over all of these studies to a meaningful risk, should we consider the reputational risk at a company level, that re-identification may be possible for any one of their studies? Or should we work at the level of the pharmaceutical industry, i.e. thinking over all companies and studies, as a demonstration attack successfully re-identifying a subject will reflect poorly on the entire industry and may cause issues in subject recruitment for all companies in their future studies? These questions will inform the levels of risk we are willing to accept and the thresholds that are adopted in quantitative assessments, which should be set in a transparent and defendable way taking into account the many factors contributing to the data situation including the sensitivity of the data and the potential impact on all stakeholders, in particular the study subjects.

The current anonymisation of clinical study datasets cannot be performed fully automatically and requires substantial resources. While it would be nice to proactively create anonymised datasets immediately after database lock, the resource for this approach is not currently justified as today most studies are never used for secondary purposes. As of end of July 2020 CSDR listed more than 3000 available studies but has received less than 600 research proposals in its 7 years of operation since 2013.

However, once anonymisation has been performed, the anonymised data no longer falls under the provision of GDPR and it becomes impossible to extract IPD from the anonymised data as the link between the two is broken. So, all subjects that did not revoke consent for secondary data analyses before data anonymisation remain included in the anonymised data and so an early anonymisaton may help maintaining subjects in the shared dataset.

For the public sharing of CSRs which are at present also partially manually redacted a suggestion has been made to simply create them from the already de-identified datasets and thus to avoid any data privacy issues. However, CSRs contain detailed case narratives of subjects who died or who had serious adverse events during the course of the trial. These detailed case narratives contain a large amount of personal information by their nature (e.g. "A 19-year-old male with type 1 diabetes developing breast cancer …") making them very prone to re-identification risks. Therefore, case narratives must be managed carefully for the risk of re-identification as they can contain many quasi-identifiers. This is often achieved by extensive redaction which impacts the utility of the narratives as their value comes from their detailed descriptions. An alternative approach would be to remove the expectations of open sharing of narratives within CSRs and sharing them in controlled, contracted, limited sharing scenarios such as Vivli after meeting the acceptable risk thresholds for sharing.

We should remain open to adopting new technologies as and when they become available, but also remain vigilant to the risk that new technologies may allow subject re-identification on data previously thought to be anonymised.

Although we have focussed on the challenges with secondary use of data, the reuse of data should be seen as an opportunity with multiple benefits: allowing external researchers and wider society to benefit from data collected in our clinical trials, demonstrating a transparent and open approach from the pharmaceutical industry with respect to the results of their clinical trials and also allowing companies to design better clinical trials and better target their therapies through internal analyses.

## Acknowledgments
The authors acknowledge the review and contributions of the PSI Data Sharing Special Interest Group, Rebecca Sudlow from Roche, Mimmi Sundler from AstraZeneca, Katherine Tucker from Roche, Lucy Frith from GlaxoSmithKline, Stef James from AstraZeneca, Hoi Shen Radcliffe from Amgen, Mahumood Hameed from Eisai, Tamsin Sargood from Johnson and Johnson, Catrin Tudor Smith from University of Liverpool, Carrol Gable from University of Liverpool, Matt Sydes from MRC Clinical Trials Unit at UCL and Michael Robling from Cardiff University.

# APPENDIX - DEFINITIONS

This appendix provides an overview of the definitions and methodology most commonly used for redaction of a CSR before making it publicly available, and for the anonymisation of IPD made available at designated secure websites like CSDR [4].

CT documents, such as the CSR, and IPD will be referred to jointly as CT data, unless otherwise specified.

As these are still early days in the development of the field of anonymisation, it's common to find inconsistencies on how terms are defined and used, for example de-identification and anonymisation are frequently being used interchangeably. To remediate this situation, we have chosen to use the following sources for the definitions below: Health Canada Guidance Document [11], EMA Guidance document [12], El Emam and Arbuckle [13] and PHUSE Data Transparency Workstream best practice guide [14], unless otherwise noted.

**Personal information** is the subject level data which can be linked to an identifiable natural person directly or indirectly, in particular by reference to details such as name, identification number, location data or to one or more factors specific to the physical, physiological, genetic, mental, economic, cultural or social identity of that natural person.

**Direct Identifiers(DI)** in a data set are those data fields that can be used to uniquely identify an individual (e.g., study participant ID, government assigned identifier like social security number, exact address, telephone number, email address, government assigned identifier) without additional information or cross-linking other information that is in the public domain.

**Quasi Identifiers or Indirect Identifiers** are the data fields which in connection with other information can be used to identify an individual with high probability (eg, postal code, date of birth, age at baseline, race, gender, medical information, events, specific findings, etc).

**Re-identification** occurs when the association between a set of identifying data and the data subject found in data/documents is re-established.

**Anonymisation** is usually defined as the overall process of modifying identifiable information in data and/or documents to reduce the risk of re-identification and to protect the privacy of data subjects, including clinical study participants by modifying their personal information in a way to satisfy:
- The data is stripped irrevocably of direct identifiers;
- Other identifiers are removed, obscured, aggregated or altered;
- The process is irreversible as it removes the association between original direct identifiers and the transformed identifiers;
- The residual risk of re-identification of a subject from the remaining indirect identifiers is assessed and controlled to a level usually below a pre-defined threshold;
- The data environment is restricted.

**De-identification** is a general term for any process of removing the association between a set of identifying data and the data subject found in data/documents. The association between data and subject is removed by suppressing, obscuring, aggregating, or altering identifiable data. De-identification deals with the transformation of quasi-identifiers (QI) by:
**Grouping** or **Aggregating**: for example, sites, countries, region, race or ethnicity, medical records, labs results may be grouped so that every single category has at least a certain number of subjects.
**Removing**: If it is not possible to use the techniques above to produce an adequate change. For example, the variable race may be removed when a subject's race can only be grouped as Whites vs. a few others.

**Pseudonymisation** is a type of de-identification that both removes the association with a data subject and adds an association between a particular set of characteristics relating to the data subject and one or more pseudonyms. Typically, pseudonymization is implemented by replacing direct identifiers (e.g, a name, a subject ID) with a pseudonym, such as a randomly generated value.

**Redaction** or masking is the separation of disclosable from non-disclosable information by blocking out individual words, sentences, or paragraphs or by removing of whole pages or sections prior to the release of the document.

For anonymised datasets, there are two main approaches to address the CT subject's risk of re-identification: quantitative and qualitative.

The **Quantitative** approach calculates the numerical risk that a subject in the anonymised datasets has of being re-identified. Two common metrics to calculate the risk are the average risk and maximum risk. The quantitative approach is the method preferred by Health Canada [11] and EMA [12], who suggest adopting a 9% re-identification risk threshold (risk=0.09).

The **Qualitative** approach is very similar to that of a quantitative approach, the difference being that the risk that any subject will be re-identified is assessed qualitatively as low, medium or high. There is no clear guidance in the literature on how to propose a robust qualitative assessment, but a defendable justification of the approach is expected.

**Data Utility** A summary term describing the value of a given data release as an analytical resource. This comprises the data's analytical completeness and its analytical validity. Disclosure control methods usually have an adverse effect on data utility. Ideally, the goal of any disclosure control regime should be to maximise data utility whilst minimizing disclosure risk. In practice disclosure control decisions are a trade-off between utility and disclosure risk [15].

Please note that altering some DI like social security number, exact address, telephone number won't for the most part affect data utility as the particular values of these variables do not affect statistical analyses.

**Confidential Business Information (CBI)** defined in [15] in respect of a person to whose business or affairs the information relates, means - subject to the regulations - business information:
a. that is not publicly available,
b. in respect of which the person has taken measures that are reasonable in the circumstances to ensure that it remains not publicly available,
c. that has actual or potential economic value to the person or their competitors because it is not publicly available, and its disclosure would result in a material financial loss to the person or a material financial gain to their competitors.

Similar definition is proposed by EMA [12] and referred as '**Commercially Confidential Information' (CCI).** It means any information contained in the clinical reports submitted to EMA by the applicant/ Marketing Authorisation Holder (MAH) which is not in the public domain or publicly available and where disclosure may undermine the legitimate economic interest of the applicant/ MAH.